\documentclass[aps,prb,twocolumn,groupedaddress,showpacs,amsmath,amssymb]{revtex4}

\usepackage{graphicx}
\usepackage{dcolumn}
\usepackage{bm}

\begin{document}

\title{Optimizing the Bi$_{2-x}$Sb$_{x}$Te$_{3-y}$Se$_{y}$ solid solutions to 
approach the intrinsic topological insulator regime}

\author{Zhi Ren}
\author{A. A. Taskin}
\author{Satoshi Sasaki}
\author{Kouji Segawa}
\author{Yoichi Ando}

\affiliation{Institute of Scientific and Industrial Research,
Osaka University, Ibaraki, Osaka 567-0047, Japan}

\date{\today}

\begin{abstract}

To optimize the bulk-insulating behavior in the topological insulator
materials having the tetradymite structure, we have synthesized and
characterized single-crystal samples of
Bi$_{2-x}$Sb$_{x}$Te$_{3-y}$Se$_{y}$ (BSTS) solid solution at various
compositions. We have elucidated that there are a series of ``intrinsic"
compositions where the acceptors and donors compensate each other and
present a maximally bulk-insulating behavior. At such compositions, the
resistivity can become as large as several $\Omega$cm at low temperature
and one can infer the role of the surface-transport channel in the
non-linear Hall effect. In particular, the composition of
Bi$_{1.5}$Sb$_{0.5}$Te$_{1.7}$Se$_{1.3}$ achieves the lowest bulk
carrier density and appears to be best suited for surface transport
studies.

\end{abstract}

\pacs{73.25.+i, 74.62.Dh, 72.20.My, 73.20.At}

%

\maketitle
\section{Introduction}

The three-dimensional (3D) topological insulator (TI)
\cite{K1,MB,Roy,K2,Qi} possesses spin-momentum-locked gapless surface
states that are expected to host a variety of interesting quantum
phenomena.\cite{Kane,ZhangSC} While a number of materials have been
identified to be 3D TIs,
\cite{H1,Taskin,Matsuda,SCZ4,H5,Shen1,Sb2Te3,Sato,Kuroda,Shen2,BTS_Hasan}
most of the known TI materials are poor insulators in the bulk, and
finding ways to achieve a bulk-insulating state is an important current
theme. \cite{Ong2010,Fisher_np,checkelsky,Butch,Analytis,Eto,RenBi2Se3}
Recently, a large bulk resistivity together with clear quantum
oscillations from the surface state was observed in a ternary
tetradymite TI material Bi$_{2}$Te$_{2}$Se (BTS).\cite{Ren,OngBTS} In
this material, the chalcogens (Te and Se) are separated into different
lattice sites, forming the ordered Te-Bi-Se-Bi-Te quintuple layers. This
ordering provides the chemical environment suitable for reducing defect
formations.\cite{Ren} As a result, the chalcogen-ordered tetradymite TI
is a promising material for approaching the intrinsic
topological-insulator regime where the bulk carriers are negligible.
Since the residual bulk carrier density in BTS was found to be of the
order of 10$^{17}$ cm$^{-3}$,\cite{Ren} it is desirable to find an
appropriated route to further improve the bulk-insulating properties of
the chalcogen-ordered tetradymite TI material.

In this respect, the tetradymite solid solution
Bi$_{2-x}$Sb$_{x}$Te$_{3-y}$Se$_{y}$ (BSTS) is interesting, because it
has been known that some special combinations of $x$ and $y$ in this
solid-solution system yield very low conductivity. \cite{Teramoto}
Except for a narrow composition region close to Sb$_{2}$Se$_{3}$, BSTS
takes the same rhombohedral structure as its three end members
Bi$_{2}$Te$_{3}$, Bi$_{2}$Se$_{3}$, and Sb$_{2}$Te$_{3}$, all are known
to be TIs.\cite{SCZ4,H5,Shen1,Sb2Te3} Hence, BSTS is naturally expected
to be a TI as long as its structure is rhombohedral. Note that the BTS
compound, where a perfect chalcogen ordering is presumably achieved, can
be regarded as a member of this family ($x$ = 0, $y$ = 1).

Originally, the BSTS solid solution attracted attention because of its
promising thermoelectric properties for near room-temperature
applications.\cite{BSTSthermoelectric} Decades of efforts were devoted
to optimizing the thermoelectric performance through tuning the
composition and/or doping in the degenerate regime, resulting in a
dimensionless figure of merit $ZT$ (= $S^{2}T/\rho\kappa$, where $S$ is
the thermopower, $\rho$ is the electrical resistivity, and $\kappa$ is
the thermal conductivity) close to 1. By contrast, nearly no effort was
made to obtain insulating behavior in this family; nevertheless, because
two prevailing types of charged defects that exist in BSTS [(Bi,Sb)/Te
anti-site defects and the Se vacancy defects] introduce carriers of
opposite signs, a careful tuning of $x$ and $y$ would, in principle,
substantially cancel the bulk carriers and allow one to maximize the
bulk insulating properties. In fact, based on a systematic study of
polycrystalline BSTS samples at room temperature, Teramoto {\it et al.}
inferred \cite{Teramoto} that a series of such ``intrinsic" compositions
exist, and that the values of $x$ and $y$ for such compositions are
linearly coupled. Unfortunately, the temperature dependence of the
resistivity was not measured in their experiment, so it was not clear to
what extent an insulating state was actually realized.

In the present work, we performed a systematic study of the transport
properties of the BSTS solid solution for a wide range of compositions
and temperature, with the aim of maximizing the insulating property. It
was found that the compositions suggested by Teramoto {\it et al.} do
not exactly correspond to the optimal compositions for insulating
behavior. Instead, we determined a series of optimized compositions,
where the relationship between $x$ and $y$ is obviously non-linear. The
BSTS crystals at the optimized compositions with $0 \le x \le 1$ were
found to present a large bulk resistivity exceeding 1 $\Omega$cm at low
temperature, and their transport properties at high temperature signify
the existence of a small activation gap. This result gives evidence for
the existence of a series of BSTS solid solutions where the electron and
hole carriers are nearly compensated. In particular, it appears that at
$(x,y) = (0.5,1.3)$ (Bi$_{1.5}$Sb$_{0.5}$Te$_{1.7}$Se$_{1.3}$) the
compensation is maximally realized. Although the chances of observing
the Shubnikov-de Haas (SdH) oscillations in BSTS samples are not very 
high, one can use the
non-linear Hall effect which is always observed at low temperature 
as a tool to infer the role of the surface-transport channel.

\section{Experimental Details}

The single crystals of Bi$_{2-x}$Sb$_{x}$Te$_{3-y}$Se$_{y}$ were grown
by melting stoichiometric amounts of high purity elements [Bi, Sb, and Te
were 6N (99.9999\%), while Se was 5N (99.999\%)] in sealed evacuated
quartz tubes at 850 $^{\circ}$C for 48 h with intermittent shaking to
ensure a homogeneity of the melt, followed by cooling slowly to 550
$^{\circ}$C and then annealing at that temperature for 4 d. The
resulting crystals have a typical domain size of up to several
centimeters and are easily cleaved along the (111) plane to reveal a
shiny surface. The crystal structure of each sample was investigated by
the x-ray diffraction (XRD) analysis with the Cu $K\alpha$ emission, 
which was performed on powders
obtained by crushing the crystals.
Lattice parameters were refined by a least-squares fitting to the XRD
data with the consideration of the zero shifts. All the samples used for
transport measurements were confirmed to be single domain by using the
x-ray Laue analysis, and they were cut into thin bar-shaped specimens with the 
typical thickness of 100 $\mu$m. The
in-plane resistivity $\rho_{xx}$ and the Hall resistivity $\rho_{yx}$ were
measured by using the standard six-probe method 
in a Quantum Design Physical Properties Measurement System
(PPMS-9) down to 1.8 K, shortly after the electrical contacts were
prepared by using room-temperature-cured silver paste.

\begin{figure}
\includegraphics*[width=8.5cm]{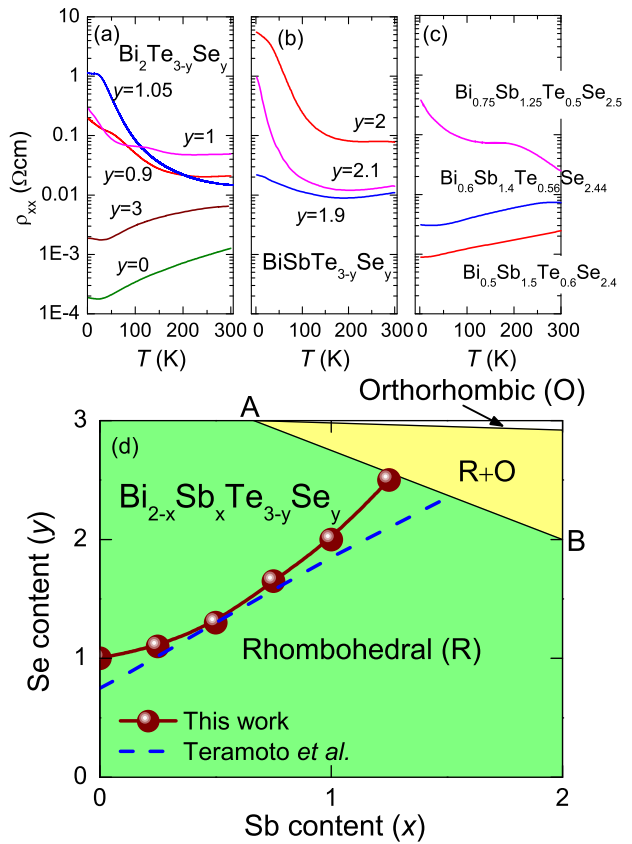}
\caption{(Color online)
(a,b) Temperature dependences of $\rho_{xx}$ for BSTS samples with 
fixed $x$ and different $y$, showing how the compositions were optimized
for the insulating behavior; panel (a) is for $x$ = 0
(Bi$_{2}$Te$_{3-y}$Se$_{y}$) and panel (b) is for $x$ = 1
(BiSbTe$_{3-y}$Se$_{y}$). (c) $\rho_{xx}(T)$ data used for optimizing
the composition along the structural phase boundary [line A-B in panel
(d)]. (d) Composition-structure diagram of the
Bi$_{2-x}$Sb$_{x}$Te$_{3-y}$Se$_{y}$ system. The circles denote the
``intrinsic" compositions determined in the present work, while the
dashed line denotes those suggested in Ref. \onlinecite{Teramoto}. 
This diagram was determined by first fixing the Sb content $x$ and 
optimizing the Se content $y$ for that $x$; the possible error 
in the optimized $y$ value for a given $x$ is the step size in the 
optimization process and is represented by the symbol size. 
}
\label{fig1}
\end{figure}

\section{Intrinsic compositions}

Figures 1(a)-(c) present how the compositions of the solid solution were
optimized for the insulating behavior. Initially, we synthesized and
characterized samples for the ``intrinsic" compositions suggested in
Ref. \onlinecite{Teramoto}. However, it turned out that, except for the
$(x,y) = (0.5,1.3)$ case, all those samples were only poorly insulating
or even metallic, with a nearly $T$-independent Hall coefficient
signifying the existence of electron carriers with the density of
10$^{18}$--10$^{19}$ cm$^{-3}$, which indicated that there are too much
Se vacancies that are not sufficiently compensated by the (Bi,Sb)/Te
anti-site defects. Therefore, we tried to optimize the composition by
increasing the Se content $y$ while fixing the Sb content $x$. The
examples for $x$ = 0 and 1 are shown in Figs. 1(a) and (b). We have also
performed similar optimization along the structural phase boundary [line
A-B in Fig. 1(d)], in which one must simultaneously vary the values of
$x$ and $y$. As shown in Fig. 1(c), the composition $(x,y) = (1.5,2.4)$
suggested in Ref. \onlinecite{Teramoto} yielded only metallic samples;
moving to $(x,y) = (1.4,2.44)$ was not sufficient, and finally at $(x,y)
= (1.25,2.5)$ we found an insulating behavior. 

The summary of the
optimized compositions are shown in the composition-structure diagram
[Fig. 1(d)], together with the linear ``intrinsic" composition line
suggested in Ref. \onlinecite{Teramoto}. The possible errors in our 
experiment, which
mostly come from the finite step size in the optimization process, are
represented by the symbol size. Obviously, the relationship between $x$
and $y$ in the true ``intrinsic" compositions is not linear but shows an
upward curvature in this diagram. It should be noted that Teramoto {\it
et al.} used polycrystalline samples and only measured the conductivity
at room temperature;\cite{Teramoto} it is therefore not surprising that
the present result does not agree with their conclusion.

\begin{figure}
\includegraphics*[width=8.5cm]{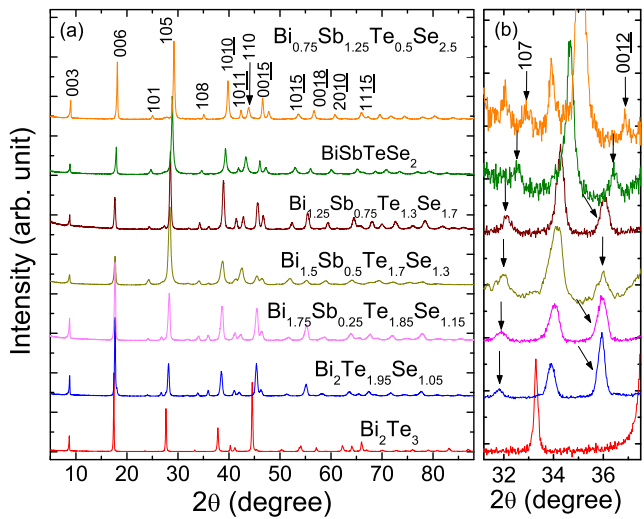}
\caption{(Color online)
(a) Power XRD patterns for the series of BSTS samples at optimized
compositions. The pattern for Bi$_{2}$Te$_{3}$ is also shown for
comparison. All the diffraction peaks can be indexed based on the
rhombohedral structure with the $R$$\bar{3}$$m$ space group. 
(b) Enlarged view of the data near the (108) peak. Arrows mark the (107)
and (00$\underline{12}$) peaks related to the ordering of the chalcogen
layers.
}
\label{fig2}
\end{figure}

\begin{table*}[btp]
\caption{Important parameters obtained for the opitimized compositions.
$\Delta$ and $\Delta^{\ast}$ are the thermal activation energies
determined from the temperature dependences of $\rho_{xx}$ and $R_{\rm
H}$, respectively. The effective acceptor concentration $N_{\rm A}^{\rm
eff}$ is estimated from the relation $R_{\rm H}^{-1} \approx e
\sqrt{N_{\rm A}^{\rm eff}N_{\rm V}} \exp(-\Delta^{\ast}/k_{\rm B}T)$,
where $N_{\rm V}$ = 5.2$\times$10$^{18}$ cm$^{-3}$ (Ref.
\onlinecite{Ren}) was assumed. Note that, among the samples with the
same composition, $\Delta$ and $\Delta^{\ast}$ vary by $\sim$10\% and $N_{\rm
A}^{\rm eff}$ could vary by a factor of three, probably due to different 
levels of defect concentrations.
}
\label{kopelTabFactors}\centering \vspace{5mm}
\begin{tabular}{cccccccccc}
\hline\hline
\\
composition& \ $a$ & \ $c$ & $\rho_{xx}^{\rm 300 K}$ & \ \   $\rho_{xx}^{\rm 1.8 K}$  & \ \  $R_{\rm H}^{\rm 300 K}$ & \ \ $R_{\rm H}^{\rm 1.8 K}$ & \ \ $\Delta$ \ & \ \ $\Delta^{\ast}$\ & $N_{\rm A}^{\rm eff}$\\
& \ \ (\AA)& \ \ (\AA)& \ \ (m$\Omega$cm) & \ \ ($\Omega$cm)   & (cm$^{3}$/C) & (cm$^{3}$/C) & \ \ \ (meV) \ \ \ & (meV) & \ \ (cm$^{-3}$) \ \ \\\hline
Bi$_{2}$Te$_{1.95}$Se$_{1.05}$ &\ $4.28$&\ \ $29.86$&\ $14$& $1.1$ & $2.6$ & $-200$ & $22$  &\ $33$ & \ \ \  9$\times$10$^{18}$ \ \ \\
Bi$_{1.75}$Sb$_{0.25}$Te$_{1.85}$Se$_{1.15}$ &\ $4.26$&\ \ $29.84$&\ $25$& $2.1$ & $2.3$ & $-610$ & $43$ &\  $60$ & \ \ \  5$\times$10$^{19}$ \ \ \\
Bi$_{1.5}$Sb$_{0.5}$Te$_{1.7}$Se$_{1.3}$&\ $4.24$&\ \ $29.83$&\ $140$& $4.5$ & $2.7$ & $-910$ & $53$  &\  $65$ & \ \ \ 6$\times$10$^{19}$ \ \ \\
Bi$_{1.25}$Sb$_{0.75}$Te$_{1.3}$Se$_{1.7}$&\ $4.20$&\ \ $29.58$&\ $44$& $2.2$ & $2.7$ & $-370$ & $38$ & \  $48$ & \ \ \  2$\times$10$^{19}$ \ \ \\
BiSbTeSe$_{2}$&\ $4.16$&\ \ $29.41$&\ $77$& $5.5$ & $2.4$ & \ $802$ & $30$ &\  $22$ & \ \ \   1$\times$10$^{18}$ \ \ \\
Bi$_{0.75}$Sb$_{1.25}$Te$_{0.5}$Se$_{2.5}$&\ $4.12$&\ \ $29.16$&\ $25$& $0.38$ & $-10.6$ &\ $207$ & \ --- & \ --- & \  --- \ \ \\
\hline\hline
\end{tabular}
\end{table*}

\section{Crystal structure}

Figure 2(a) shows the powder XRD patterns for the series of BSTS solid
solutions at the optimized compositions, together with the
Bi$_{2}$Te$_{3}$ data which is shown for comparison. The patterns for
all the samples are essentially the same and can be well indexed with
the rhombohedral structure (space group $R\bar{3}m$), confirming that
the solid solutions maintain the same crystal structure as their binary
end members Bi$_{2}$Te$_{3}$, Bi$_{2}$Se$_{3}$, and Sb$_{2}$Te$_{3}$. The
refined lattice parameters are summarized in Table I. As the system
moves toward Sb$_{2}$Se$_{3}$, both the $a$ and $c$ lattice parameters tend to
decease, which is reasonable because the ionic radius of Sb (Se) is
smaller that that of Bi (Te).

It is important to note that the (107) and (00$\underline{12}$) peaks,
both of which are characteristic of the compounds with the BTS-type
chalcogen ordering, are present in all the BSTS samples, as can seen in
the magnified plot [Fig. 2(b)]. This indicates that the ordering of the
chalcogen layers is preserved to some extent despite a large change in
stoichiometry. To understand this observation, it is useful to recall
the structure of the tetradymite, which is composed of stacked A$^{\rm
V}_{2}$B$^{\rm VI}_{3}$ (A$^{\rm V}$ = Bi, Sb; B$^{\rm VI}$ = Te, Se)
quintuple layers. Within the quintuple layer, the atoms are arranged in
the sequence of B$^{\rm VI}(1)-$A$^{\rm V}-$B$^{\rm VI}(2)-$A$^{\rm
V}-$B$^{\rm VI}(1)$, where B$^{\rm VI}$(1) and B$^{\rm VI}$(2) are two
inequivalent crystallographic sites. Because Se has a larger
electronegativity than Te, the Se atoms preferentially occupy the
B$^{\rm VI}$(2) site and the remaining $(y-1)\rm{Se}+(3-y){\rm Te}$
atoms are distributed randomly in the B$^{\rm VI}$(1) site when $y$ is
larger than 1.\cite{BSTSordering} Such an asymmetric arrangement of the
chalcogen layers bears similarity to that in BTS and accounts for the
XRD results.

\begin{figure}\includegraphics*[width=7.5cm]{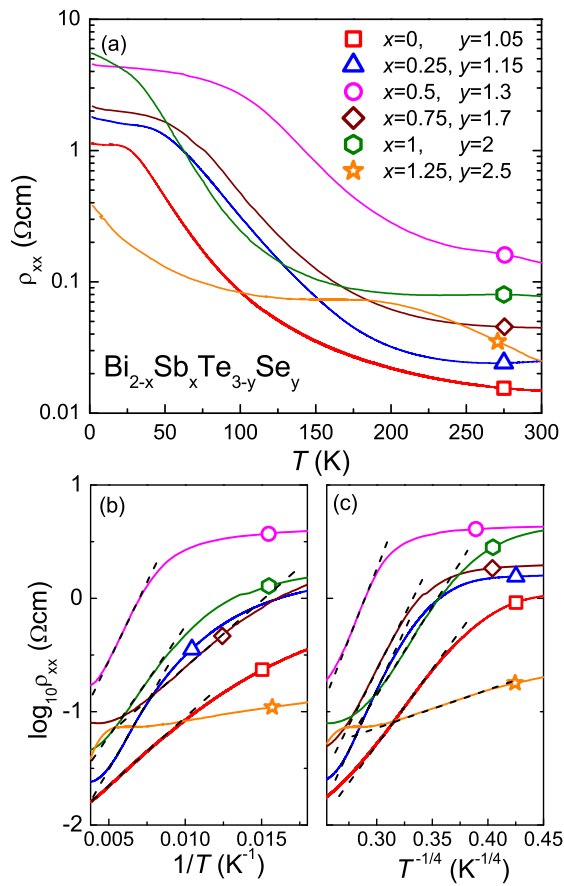}
\caption{(Color online)
(a) Temperature dependences of $\rho_{xx}$ for the series of BSTS
samples at optimized compositions. Note that the vertical axis is in the
logarithmic scale with base 10. 
(b) and (c) are the Arrhenius plot [$\log_{10} \rho_{xx}$ vs $1/T$] and
the 3D-VRH plot [$\log_{10} \rho_{xx}$ vs $T^{-1/4}$] of the $\rho_{xx}(T)$
data, respectively. Dashed lines represent linear fittings to the regions
where the activation and the VRH behaviors can be recognized.
}
\label{fig3}
\end{figure}

\section{Resistivity}

Figure 3(a) shows the typical temperature dependences of $\rho_{xx}$ for
the optimized compositions. It should be noted that the $\rho_{xx}$
value was found to be sample dependent and could vary by a factor of
three within the same composition, probably due to different levels of
defect concentrations. Nevertheless, the activated behavior was
essentially reproducible within the same batch and the variation in
terms of the activation energy was about 10\%. The magnitude of
$\rho_{xx}$ at room temperature in Fig. 3(a) ranges from 25 to 140 m$\Omega$cm, which
is generally larger than that found in BTS.\cite{Ren} Except for the
$(x,y) = (1.25,2.5)$ composition which has the largest Se content, the
overall temperature dependence of $\rho_{xx}$ is similar and can be
divided into three regimes:

(i) {\it Activated regime:} 
In the temperature range above $\sim$100 K, the $\rho_{xx}(T)$ data can be
fitted with the Arrhenius law 
\begin{equation}
\rho_{xx} \sim \exp(\Delta/k_{\rm B}T), 
\end{equation}
where $\Delta$ is the activation energy and $k_{\rm B}$ is the Boltzmann
constant. In Fig. 3(b), we show the Arrhenius plot of the data together
with linear fittings to obtain $\Delta$. The $\Delta$ values are summarized in
Table I, and they vary from 22 to 53 meV; however, since the
temperature range of this activated behavior is not very wide, the
obtained $\Delta$ should not be taken too literally.

(ii) {\it Variable-range hopping regime:}
Below the temperature range of the activated behavior mentioned above, 
the $\rho_{xx}(T)$ appears to be better described by the 3D 
variable-range hopping (VRH) behavior,
\begin{equation}
\rho_{xx} \sim \rm exp[(\it T/T_{\rm 0})^{\rm -1/4}],
\end{equation}
where $T_{\rm 0}$ is a constant which depends on the density of states
at the Fermi level $E_{\rm F}$. For example, in the data for
Bi$_{2}$Te$_{1.95}$Se$_{1.05}$ (marked by square in Fig. 3), the VRH
behavior holds for 60 -- 140 K, while the activated behavior holds for
110 -- 300 K. However, it should be noted that the temperature range
where this VRH behavior appears to hold can have a significant overlap
with the activated temperature range. For example, in the case of
Bi$_{1.25}$Sb$_{0.75}$Te$_{1.3}$Se$_{1.7}$ (marked by diamond in Fig. 3)
where the distinction is the most ambiguous, the activation behavior and
the VRH behavior appear to hold in similar ranges, 85 -- 140 K and 90 --
170 K, respectively. Hence, the distinction between the two transport
mechanisms can be a subtle issue, and it probably depends on the level
of disorder in the sample.

(iii) {\it Saturation regime:}
At low temperature, $\rho_{xx}$ tends to saturate, rather than to
diverge as expected for intrinsic semiconductors, implying the existence
of some extended states at $E_{\rm F}$ in the zero-temperature limit. In
this regime, the magnetic-field dependence of the Hall resistivity
$\rho_{yx}$ is generally non-linear (as will be described in detail
later), which points to the existence of two or more transport channels.
In the BTS compound, it was elucidated \cite{Ren} that the topological
surface state and a degenerate bulk impurity band both contribute to the
low-temperature saturated resistivity. In addition, our recent study of
the Bi$_{1.5}$Sb$_{0.5}$Te$_{1.7}$Se$_{1.3}$ compound has identified
\cite{BSTStaskin} the third contribution of the accumulation layer (which is
caused by the surface band bending and is topologically trivial) in
samples where the Hall coefficient is positive at low temperature.

\begin{figure}\includegraphics*[width=8.5cm]{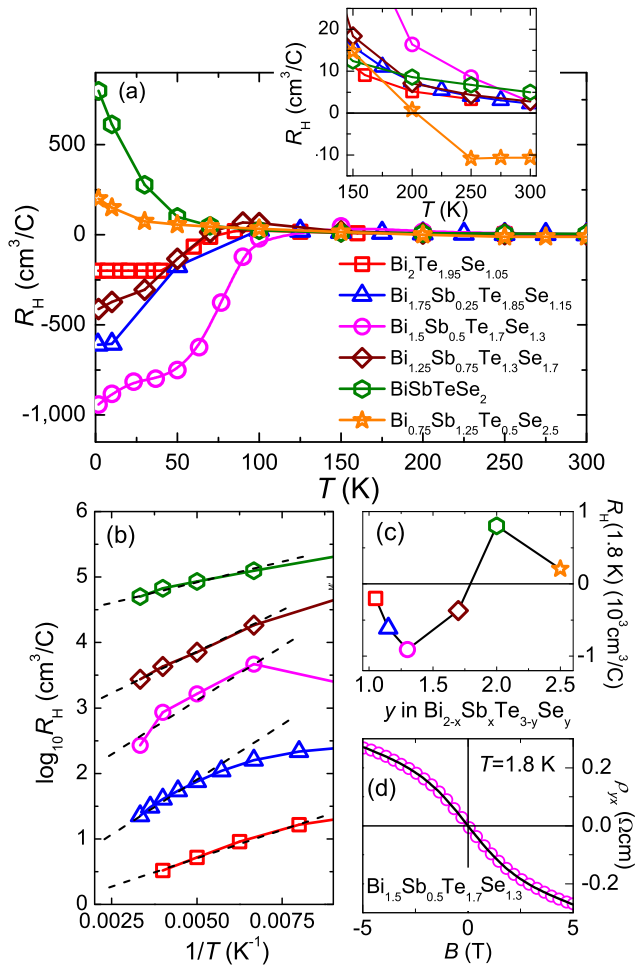}
\caption{(Color online)
(a) Temperature dependences of the low-field Hall coefficient $R_{\rm
H}$ for the series of BSTS samples at optimized compositions. Inset
shows an enlarged view of the data between 150 and 300 K. (b) Arrhenius
plot of $R_{\rm H}(T)$ at high temperature for the series of samples
except Bi$_{0.75}$Sb$_{1.25}$Te$_{0.5}$Se$_{2.5}$ where the activation
behavior was not clearly observed; the data are shifted vertically for
clarity. Dashed lines are the Arrhenius-law fittings to extract the
activation energy $\Delta^*$. (c) The value of $R_{\rm H}$ at 1.8 K
plotted as a function of the Se content $y$. (d) Magnetic-field
dependence of $\rho_{yx}$ for the
Bi$_{1.5}$Sb$_{0.5}$Te$_{1.7}$Se$_{1.3}$ sample at 1.8 K; solid line is
a two-band-model fitting (Ref. \onlinecite{Ren}) to the data. The
obtained fitting parameters give the surface conductance of
1.4$\times$10$^{-4}$ $\Omega^{-1}$ and the bulk conductivity of 0.21
$\Omega^{-1}$cm$^{-1}$; since the measured sample was 120 $\mu$m thick,
the surface contribution was 5\% of the total conductance.
}
\label{fig4}
\end{figure}

\section{Hall coefficient}

To further characterize the transport properties, we have performed the
Hall measurements. Figure 4(a) shows the temperature dependences of the
low-field Hall coefficient $R_{\rm H}$, defined as $R_{\rm H} =
\rho_{yx}/B$ near $B$ = 0, for the same series of samples. Above
$\sim$150 K, $\rho_{yx}$ is always linear in $B$ so that $\rho_{yx}/B$ 
is magnetic-field independent, and the resulting
$R_{\rm H}$ is positive and shows a thermally-activation behavior in all
the BSTS samples except Bi$_{0.75}$Sb$_{1.25}$Te$_{0.5}$Se$_{2.5}$. The
positive and activated $R_{\rm H}(T)$ behavior indicates that the
dominant charge carriers are thermally-excited holes in the bulk valence
band. By contrast, the high-temperature $R_{\rm H}$ in
Bi$_{0.75}$Sb$_{1.25}$Te$_{0.5}$Se$_{2.5}$ is negative and only weakly
dependent on temperature, which indicates that the electron carriers
created by Se vacancies in this compound 
are degenerate near room temperature. Thus, the
sign change in the high-temperature $R_{\rm H}$ at the largest Se
content $y$ in our series of samples reflects a change in the dominant
defects from the Bi(Sb)/Te anti-sites (acceptors) to the Se vacancies
(donors). Note that in those samples both acceptors and donors are
expected to exist, and the observed charge-carrier concentration is
likely to be determined by their competition.

From the Arrhenius plot of $R_{\rm H}(T)$ [Fig. 4(b)], one obtains the
effective acceptor concentration $N_{\rm A}^{\rm eff}$ together with the
effective activation energy $\Delta^{\ast}$, both are summarized in
Table I. It should be noted that, among the samples with the same
composition, the variation in $\Delta^{\ast}$ was within $\sim$10\%,
while $N_{\rm A}^{\rm eff}$ could vary up to a factor of three,
reflecting the variation in the resistivity behavior within the same
batch mentioned in Sec. V. One can see that the $\Delta^{\ast}$ values
are not much different from the $\Delta$ values derived from the
$\rho_{xx}(T)$ data for the same samples, and the difference is less
than 30\%. This small differences can probably be ascribed to the
temperature dependence of the mobility. Intriguingly, $N_{\rm A}^{\rm
eff}$ and $\Delta^{\ast}$ seem to be correlated; namely, larger $N_{\rm
A}^{\rm eff}$ is accompanied with larger $\Delta^{\ast}$. However, the
origin of this behavior is not clear at the moment.

At lower temperature below 100 K, $\rho_{yx}(B)$ is no longer a linear 
function of $B$ and $R_{\rm H}$ becomes very much
dependent on composition. Figure 4(c) shows this variation in terms of
the $R_{\rm H}$ value at 1.8 K plotted vs $y$.\cite{note} The largest
$|R_{\rm H}|$ was observed in Bi$_{1.5}$Sb$_{0.5}$Te$_{1.7}$Se$_{1.3}$,
where $R_{\rm H}$ = $-910$ cm$^{3}$/C at 1.8 K. This value would naively
correspond to a very low carrier density of 7$\times$10$^{15}$
cm$^{-3}$; however, as we have demonstrated for BTS, \cite{Ren} such a
naive estimate based on the low-field $R_{\rm H}$ is not reliable when there
are multiple transport channels that cause $\rho_{yx}(B)$ to become non-linear.
Indeed, $\rho_{yx}(B)$ was found to be non-linear in 
Bi$_{1.5}$Sb$_{0.5}$Te$_{1.7}$Se$_{1.3}$ [Fig. 4(d)], and an accurate
estimate of the bulk and surface carrier densities is difficult without
employing additional information from SdH
oscillations.\cite{Ren,BiSb} Nevertheless, by using the simple two-band
model to consider both the bulk and surface transport channels \cite{Ren}
(note that the accumulation layer \cite{BSTStaskin} is irrelevant here 
since the slope of $\rho_{yx}(B)$ is negative),
one can make a reasonable fitting to the data [solid line in Fig. 4(d)],
which yields the bulk carrier density 1.8$\times$10$^{16}$ cm$^{-3}$,  
the bulk mobility 73 cm$^2$/Vs, the 
surface carrier density 1.5$\times$10$^{11}$ cm$^{-2}$, and the surface mobility
2900 cm$^2$/Vs. (In the fitting procedure, the parameters were constrained 
by the measured $\rho_{xx}$ value in 0 T.) Based on this 
fitting result one can estimate the contribution of the surface transport
to the total conductance to be $\sim$5\%, which is 
reasonably large considering the thickness (120 $\mu$m)
of the measured sample.

\section{Discussions}

For the transport studies of the topological surface states, it is
desirable that their SdH oscillations are observed. In the present
series of BSTS samples, we did not observe clear SdH oscillations below
9 T. Most likely, whether the SdH oscillations can be observed depends
crucially on the homogeneity and the value of the surface carrier
density $n_{s}$,\cite{note_SdH} which may vary between samples and, even
in the same single crystal, between different cleaves. The chances of
observing the SdH oscillations were roughly 10\% in our BSTS samples. On
such lucky occasions, we were able to see both Dirac holes and electrons
on the surface of Bi$_{1.5}$Sb$_{0.5}$Te$_{1.7}$Se$_{1.3}$ samples, and
the result was reported in Ref. \onlinecite{BSTStaskin}. It should be
emphasized, however, that even though the SdH oscillations are not
always observed, one can achieve a surface-dominated transport by making
sufficiently thin samples of BSTS.\cite{BSTStaskin}

As we have already discussed in Sec. VI, the low-field $R_{\rm H}$ value
in BSTS does not simply reflect the bulk carrier density, but
it is also influenced by the contribution of the surface carriers which
presumably have a higher mobility than the bulk carriers. In view of
this situation, the non-trivial composition dependence of $R_{\rm H}$
observed at 1.8 K [Fig. 4(c)] can be interpreted in two ways: One
possibility is that it may signify the sign change of the surface
carriers as the Se concentration $y$ is increased. This is possible
because the surface band structure of BTS is similar \cite{BTS_Hasan} to that of
Bi$_2$Te$_3$ where the Dirac point is located below the top of the
valence band,\cite{Shen1} whereas with increasing $y$ the surface band structure is
expected to approach that of Bi$_2$Se$_3$ where the Dirac point is
isolated in the bulk gap;\cite{H5} therefore, if the chemical potential is pinned
to the impurity band located just above the bulk valence band,\cite{Ren}
the nature of the surface carriers could change from Dirac electrons to
Dirac holes. The other possibility is that the change in $R_{\rm H}$ is
a reflection of the change in the nature of the degenerate bulk carriers
in the impurity band. This is also possible because at low $y$ the
chemical potential is likely to be located below the center of the
acceptor impurity band, whereas with increasing $y$ the increasing
amount of donors (due to Se vacancies) would cause the acceptor impurity
band to be gradually filled and move the chemical potential to above the
center of the impurity band. To further address this issue, systematic
studies employing the SdH oscillations would be necessary.

\section{Conclusion}

In this work, we have elucidated in the Bi$_{2-x}$Sb$_{x}$Te$_{3-y}$Se$_{y}$ 
(BSTS) solid solution the existence of the ``intrinsic"
compositions where the acceptors due to the (Bi,Sb)/Te anti-site defects
and the donors due to the Se vacancy defects compensate each other and
realize a maximally bulk-insulating state. The powder XRD patterns
suggest that at those compositions optimized for the insulating
behavior, the crystal structure has an ordering of the chalcogen layers
in which the Se atoms preferentially occupy the middle of the quintuple
layer and the remaining Se and Te atoms randomly occupy the outer-most
layers. Except for Bi$_{0.75}$Sb$_{1.25}$Te$_{0.5}$Se$_{2.5}$ which is
close to the structural instability, all the BSTS samples at the
optimized compositions show large $\rho_{xx}$ values exceeding 1 $\Omega$cm at
low temperature, together with an activated behavior at high temperature
signifying the existence of an activation gap for the bulk carriers. The
$B$ dependence of $\rho_{yx}$ at low temperature points to the role of
the surface transport channel, and the non-trivial composition
dependence of the low-field $R_{\rm H}$ reflects either the change in
the surface band structure or the change in the bulk carriers in the
impurity band. The $|R_{\rm H}|$ value at 1.8 K was found to be the
largest in Bi$_{1.5}$Sb$_{0.5}$Te$_{1.7}$Se$_{1.3}$ and this
composition appears to have achieved the lowest bulk carrier density.

\begin{acknowledgments}
This work was supported by JSPS (NEXT Program), MEXT (Innovative Area
``Topological Quantum Phenomena" KAKENHI 22103004), and AFOSR (AOARD
10-4103).
\end{acknowledgments}


\begin{thebibliography}{99}
\bibitem{note}
In our recent study of Bi$_{1.5}$Sb$_{0.5}$Te$_{1.7}$Se$_{1.3}$, we
observed that in very thin samples the low-temperature $R_{\rm H}$ can
be positive immediately after the cleavage, and it changes sign with
time (Ref. \onlinecite{BSTStaskin}). All the samples measured in the
present work were relatively thick ($\sim$100 $\mu$m) and did not show
clear time evolution in $R_{\rm H}$.
\bibitem{note_SdH}
When $n_{s}$ is inhomogeneous, the SdH oscillation frequency varies across
the sample and the oscillations are smeared; also, when $n_{s}$ is too
small, the SdH oscillation frequency becomes too low to be detected in our
limited magnetic-field range.
\end{thebibliography}
\end{document}